\documentstyle[12pt]{article}

\font\blackboard=msbm10 at 12pt
\font\blackboards=msbm7
\font\blackboardss=msbm5
\newfam\black
\textfont\black=\blackboard
\scriptfont\black=\blackboards
\scriptscriptfont\black=\blackboardss

\newcommand{\ba}{\begin{array}}
\newcommand{\ea}{\end{array}}
\newcommand{\be}{\begin{equation}}
\newcommand{\ee}{\end{equation}}
\newcommand{\bea}{\begin{eqnarray}}
\newcommand{\eea}{\end{eqnarray}}
\newcommand{\beas}{\begin{eqnarray*}}
\newcommand{\eeas}{\end{eqnarray*}}

\def\half{{1 \over 2}}

\def\laplace{{\kern1pt\vbox{\hrule height 1.2pt\hbox{\vrule width 1.2pt\hskip
  3pt\vbox{\vskip 6pt}\hskip 3pt\vrule width 0.6pt}\hrule height 0.6pt}
  \kern1pt}}
\def\scriptlap{{\kern1pt\vbox{\hrule height 0.8pt\hbox{\vrule width 0.8pt
  \hskip2pt\vbox{\vskip 4pt}\hskip 2pt\vrule width 0.4pt}\hrule height 0.4pt}
  \kern1pt}}

\def\roughly#1{\raise.3ex\hbox{$#1$\kern-.75em\lower1ex\hbox{$\sim$}}}

\def\str{{\rm STr} \,}

\def\itt{{\cal  T}}
\def\ijj{{\cal J}}
\def\imm{{\cal M}}

\def\dx{{\dot{\bf X}}}

\newcommand{\PRL}{{\em Phys.\ Rev.\ Lett.\ }}

\newcommand{\gone}[1]{}
\begin{document}
\pagestyle{plain}
\setcounter{page}{1}

\baselineskip16pt

\begin{titlepage}

\begin{flushright}
PUPT-1769\\
hep-th/9803003
\end{flushright}
\vspace{16 mm}

\begin{center}

{\Large \bf Conservation of Supergravity Currents\\
from Matrix Theory}

\vspace{3mm}

\end{center}

\vspace{8 mm}

\begin{center}

Mark Van Raamsdonk

\vspace{3mm}
{\small \sl Department of Physics} \\
{\small \sl Joseph Henry Laboratories} \\
{\small \sl Princeton University} \\
{\small \sl Princeton, New Jersey 08544, U.S.A.} \\
{\small \tt mav @princeton.edu}

\end{center}

\vspace{1cm}

\begin{abstract}
In recent work by Kabat and Taylor, certain Matrix theory quantities have been 
identified with the spatial moments of the supergravity stress-energy tensor, 
membrane current, and fivebrane current. In this note, we determine the 
relations between these moments required by current conservation, and prove 
that these relations hold as exact Matrix Theory identities at finite N. This 
establishes conservation of the effective supergravity currents (averaged over 
the compact circle). In addition, the constraints of current conservation allow us 
to deduce Matrix theory quantities corresponding to moments of the spatial 
current of the longitudinal fivebrane charge, not previously identified.
\end{abstract}

\vspace{1cm}
\begin{flushleft}
December 1997
\end{flushleft}
\end{titlepage}
\newpage


\section{Introduction}
\label{Intro}
Since the original BFSS conjecture \cite{BFSS}, a remarkable correspondence 
between matrix quantum mechanics and eleven-dimensional supergravity 
has emerged. Matrix theory states corresponding to gravitons, membranes, and 
fivebranes have been identified, and many aspects of the interaction between 
these objects calculated from Matrix theory have been shown to agree with 
supergravity predictions (\cite{BanksReview,Susskind-review,Wati-review} and 
refs. therein).  Certain apparent discrepancies have also been pointed out, 
however (see, for example\cite{Dine-Rajaraman,Douglas-Ooguri}). 

Recently, more general results involving the interaction potentials of arbitrary 
separated objects have been proven \cite{Dan-Wati, Mark-Wati,Dan-Wati2}. In 
particular, Kabat and Taylor  \cite{Dan-Wati2} have shown that in a Matrix theory 
background corresponding 
to two arbitrary well separated objects, a series of terms in the one-loop matrix 
theory potential exactly reproduces the linearized supergravity potential arising 
from the exchange of quanta with zero longitudinal momentum. Central to this 
analysis was a proposal for the identification of certain Matrix theory quantities 
with the spatial moments of the stress-energy tensor, membrane current, and 
fivebrane current of the related supergravity theory.  These definitions give a 
precise correspondence between a given Matrix theory background and the 
distribution of matter and charges in the transverse spatial dimensions of the 
supergravity background.  Recently, this correspondence has been used to show 
that the statistical calculation of black hole entropy in the Matrix description of 
M-theory reduces to the original Gibbons-Hawking calculation of thermodynamic 
entropy \cite{Lowe}. 

Crucial to the consistency of eleven-dimensional supergravity is the 
conservation of the stress-energy tensor, membrane current, and fivebrane 
current, since these are the currents which couple to the massless particles of 
the theory. If Matrix theory actually does contain a description of supergravity, 
we expect that this current conservation should be derivable directly from matrix 
theory. Thus, an obvious consistency check for the identifications in 
\cite{Dan-Wati2} is to 
determine whether or not the moments, as defined in Matrix theory, obey the 
relations imposed by conservation of the respective currents.  This is the aim of 
the present paper.

In section 2,  the relations between the moments of a spacetime current dictated 
by current conservation are derived. In section 3, we show that all such relations 
involving the moments identified in \cite{Dan-Wati2} from Matrix theory are 
satisfied as 
precise, finite-N Matrix theory identities.  This demonstrates that the definitions 
given are consistent with supergravity current conservation. Alternatively, our 
results prove the conservation of ten-dimensional spacetime currents which we 
associate with the supergravity stress energy tensor and membrane current 
averaged over the compact circle. In the case of the fivebrane current, previous 
work suggested only a definition for the static longitudinal fivebrane charge, 
but the relations we derive give a natural Matrix theory definition for moments 
of the spatial current associated with this charge. Section 4 contains a 
discussion of our results.

\newpage

\section{Predictions of conservation equations}

In this section, we wish to derive the relations among the moments of a given 
spacetime current imposed by current conservation.

We consider a current $C^{\mu}(x)$ with compact support in a (for now) 
uncompactified spacetime with spatial directions $i=1,...,d$. Define the spatial 
moments of this current as
\[
{\cal C}^{\mu (l_1\cdots l_n)} = \int d^d x [C^{\mu}(x)x^{l_1}\cdots x^{l_n}]
\]
Now, assuming the current is conserved, \footnote{In this work we will always be 
dealing with expressions appropriate to linearized gravity with a flat 
background, thus it is not necessary to use a covariant derivative.}
\[
\partial_{\mu}C^{\mu}(x) = 0 
\]
we have

\beas
0 &=& \int d^d x [\partial_{\mu}C^{\mu}x^{l_1}\cdots x^{l_n}]\\
 &=& \int d^d x [(\partial_t C^t + \partial_i C^i)x^{l_1}\cdots x^{l_n}]
\eeas
so, integrating by parts and using the definition above, we find
\be
\partial_t{\cal C}^{t(l_1\cdots l_n)} = {\cal C}^{l_1(l_2\cdots l_n)} + \cdots + {\cal 
C}^{l_n 
(l_1\cdots l_{n-1})}
\label{moment_eqns}
\ee
This is our desired relation. It expressed the time derivative of the moments of a 
conserved charge in terms of moments of the associated spatial current.

   In the case of a theory with a compact circular direction $x^-$, relevant to finite 
N Matrix theory, $x^-$ is not continuous around the circle, so we should define 
moments with a periodic variable in the integral (e.g. Fourier modes around the 
circle). Thus, we take
 \be
{\cal C}_m^{\mu (l_1\cdots l_n)} = \int dx^- \int d^d x 
[C^{\mu}(x)x^{l_1}\cdots x^{l_n}e^{imx^-/R}]
\label{fourier}
\ee
In this case, the relevant relations become:
\[
\partial_+{\cal C}_m^{+(l_1\cdots l_n)} = {\cal C}_m^{l_1(l_2\cdots l_n)} + \cdots 
+ {\cal C}_m^{l_n (l_1\cdots l_{n-1})} + {im \over R} {\cal C}_m^{-(l_1\cdots l_n)}
\]
where we have use indices appropriate to a lightlike compactified theory.
The relations for $m=0$,
\be
\partial_+{\cal C}_0^{+(l_1\cdots l_n)} = {\cal C}_0^{l_1(l_2\cdots l_n)} + \cdots 
+ {\cal 
C}_0^{l_n (l_1\cdots l_{n-1})} 
\label{zero_eqns},
\ee
simply express the fact that the averaged current $\int 
dx^- C^{\mu}(x),$ $\mu \in \{+,1,...,9\}$ obeys the conservation equations for 
uncompactified ten-dimensional spacetime (compare (\ref{moment_eqns})). 

   In fact, it is these $m=0$ moments, of the stress-energy tensor, membrane 
current and fivebrane current, that appear in the supergravity potential between 
widely separated objects arising from the exchange of quanta with zero 
longitudinal 
momentum.  Hence, it is these objects for which corresponding Matrix theory 
quantities were deduced in \cite{Dan-Wati2}.  In the following sections, we recall 
these 
identifications and show that these do in fact satisfy all relations 
(\ref{zero_eqns}) implied by 
current conservation. 
 
\section{Proof of conservation relations}

In this section, we review the identifications made in \cite{Dan-Wati2} between 
Matrix theory 
quantities and moments of the supergravity stress-energy tensor, membrane 
current, and fivebrane current, and show that these objects satisfy all relations 
required by conservation of the respective supergravity currents. 
   
\subsection{Stress-Energy Tensor}

In \cite{Dan-Wati2}, the following Matrix theory quantities were identified as the 
transverse spatial moments of the stress-energy tensor (integrated around the 
compact circle) for a given Matrix theory background ${\bf X}^i$ satisfying the 
equations of motion. Defining $F^{ij} = i[{\bf X}^i,{\bf X}^j]$, we have (supressing 
the subscript $0$ which denotes the zeroeth fourier mode)

\beas
\label{eq:MatrixT}
\itt^{--(l_1\cdots l_n)} & = & {1 \over R} \; \str \left(({1 \over 4} \dx^i \dx^i \dx^j 
\dx^j 
+ {1 \over 4} \dx^i \dx^i F^{jk}F^{jk} + \dx^i \dx^j F^{ik} F^{kj} \right.  
\\
& & \hspace{1.2in} \left. + {1 \over 4} F^{ij} F^{jk} F^{kl}F^{li} - {1 \over 16} F^{ij} 
F^{ij} F^{kl} F^{kl}){\bf X}^{l_1}\cdots{\bf X}^{l_n} \right) \\
\itt^{-i(l_1\cdots l_n)} & = & {1 \over R} \;\str \left((\half \dx^i \dx^j \dx^j +
{1 \over 4} \dx^i F^{jk} F^{jk} 
                  + F^{ij} F^{jk} \dx^k ){\bf X}^{l_1}\cdots{\bf X}^{l_n}\right)  
\\
\itt^{+-(l_1\cdots l_n)} & = & {1 \over R} \;\str \left((\half \dx^i \dx^i + {1
\over 4} F^{ij} F^{ij} ){\bf X}^{l_1}\cdots{\bf X}^{l_n}\right)  \\ 
\itt^{ij(l_1\cdots l_n)} & = & {1 \over R}  \;\str \left( (\dx^i \dx^j + F^{ik}
F^{kj}){\bf X}^{l_1}\cdots{\bf X}^{l_n} \right)  \\ 
\itt^{+i(l_1\cdots l_n)} & = & {1 \over R} \;\str \left(\dx^i {\bf X}^{l_1}\cdots{\bf 
X}^{l_n}\right) \\
\itt^{++(l_1\cdots l_n)} & = & {1 \over R} \;\str \left({\bf X}^{l_1}\cdots{\bf 
X}^{l_n}\right) 
\eeas
Here, $\str$ is a symmetrized trace, in which for each term we take an average of 
all possible orderings of the independently written factors (i.e. $F^{ij}$ is to be 
treated as a unit).
   From equation (\ref{zero_eqns}) we find that current conservation implies the 
following 
relations among these moments
\[
\partial_t\itt^{+\mu(l_1\cdots l_n)} = \itt^{\mu l_1(l_2\cdots l_n)} + \cdots + 
\itt^{\mu l_n (l_1\cdots l_{n-1})}
\]
for arbitrary $n$ and $\mu \in \{+,-,i\}$. (Recall that $x^+$ is the Matrix theory 
time). 
We now prove these as exact Matrix theory 
identities. For $\mu = +$, we have (using a bar to denote an omitted index)
\beas
\sum_{j=1}^n\itt^{+ l_j (l_1\cdots {\bar l}_j\cdots l_n)} &=& {1 \over R} 
\sum_{j=1}^n\;\str\left({\bf X}^{l_1} \cdots {\bf X}^{l_{j-1}}\dx^{l_j}{\bf 
X}^{l_{j+1}} \cdots {\bf X}^{l_n}\right)\\
&=& {1 \over R} \partial_t \;\str({\bf X}^{l_1}\cdots {\bf X}^{l_n})\\
&=& \partial_t\itt^{++(l_1\cdots l_n)} 
\eeas
In the next case ($\mu = i$), using equation (\ref{id1}) from the Appendix, we 
have
\beas
\sum_{j=1}^n\itt^{i l_j (l_1\cdots {\bar l}_j\cdots l_n)} &=& {1 \over R} 
\sum_{j=1}^n\;\str\left(\dx^i{\bf X}^{l_1}\cdots {\bf X}^{l_{j-1}}\dx^{l_j}{\bf 
X}^{l_{j+1}}\cdots {\bf X}^{l_n} \right. \\
& & \hspace{1in} \left. - [{\bf 
X}^i,{\bf X}^k]{\bf X}^{l_1}\cdots{\bf X}^{l_{j-1}}[{\bf X}^k,{\bf X}^{l_j}]{\bf 
X}^{l_{j+1}} \cdots {\bf 
X}^{l_n}\right)\\
&=& {1 \over R} \;\str \left(\dx^i \partial_t({\bf X}^{l_1}\cdots {\bf X}^{l_n}) - 
[[{\bf X}^i,{\bf X}^k],{\bf X}^k]{\bf X}^{l_1}\cdots {\bf X}^{l_n}\right)\\
&=& {1 \over R} \;\str (\dx^i \partial_t({\bf X}^{l_1}\cdots {\bf X}^{l_n}) + 
\ddot{\bf X}^i{\bf X}^{l_1}\cdots {\bf X}^{l_n} )\\
&=& {1 \over R} \partial_t \;\str(\dx^i{\bf X}^{l_1}\cdots {\bf X}^{l_n})\\
&=& \partial_t\itt^{+i(l_1\cdots l_n)}
\eeas
where we have used the equations of motion $\ddot{\bf X}^i = - [[{\bf X}^i,{\bf 
X}^k],{\bf X}^k]$ in the third line. We emphasize that each commutator is to be 
treated as a unit in the symmetrization, throughout this work. For the last case 
($\mu = -$), using equation 
(\ref{id2}) 
from the appendix, we have:
\beas
\sum_{j=1}^n\itt^{-l_j (l_1\cdots {\bar l}_j\cdots l_n)} &=& {1 \over R} 
\sum_{j=1}^n \;\str\left(({1 
\over 
2}\dx^i\dx^i+{1 \over 4}F^{kl}F^{kl}){\bf X}^{l_1}\cdots{\bf X}^{l_{j-1}} 
\dx^{l_j}{\bf X}^{l_{j+1}}\cdots {\bf 
X}^{l_n} \right.\\
& &\hspace{1.5in} \left. - \dx^i[{\bf X}^i,{\bf X}^k]{\bf X}^{l_1}\cdots{\bf 
X}^{l_{j-1}}[{\bf 
X}^k,{\bf X}^{l_j}]{\bf X}^{l_{j+1}}\cdots {\bf X}^{l_n}\right)\\
&=& {1 \over R} \;\str\left(({1 \over 2}\dx^i\dx^i+{1 \over 
4}F^{kl}F^{kl})\partial_t({\bf X}^{l_1}\cdots{\bf X}^{l_n}) \right.\\
& &\hspace{1.5in} \left. - (\dx^i[[{\bf X}^i,{\bf X}^k],{\bf X}^k] + [\dx^i,{\bf 
X}^k][{\bf X}^i,{\bf X}^k]){\bf X}^{l_1}\cdots{\bf X}^{l_n}\right)\\
&=& {1 \over R} \;\str\left(({1 \over 2}\dx^i\dx^i+{1 \over 
4}F^{kl}F^{kl})\partial_t({\bf X}^{l_1}\cdots{\bf X}^{l_n}) \right.\\
& &\hspace{1.5in} \left. + (\dx^i\ddot{\bf X}^i + 
{1 
\over 2} \dot{F}^{kl}F^{kl}){\bf X}^{l_1}\cdots{\bf X}^{l_n}\right)\\
&=&{1 \over R} \partial_t \;\str\left(({1 \over 2}\dx^i\dx^i+{1 \over 
4}F^{kl}F^{kl}){\bf X}^{l_1}\cdots{\bf X}^{l_n}\right)\\
&=&\partial_t\itt^{+-(l_1\cdots l_n)},
\eeas
as desired.

Therefore, the Matrix theory quantities $\itt^{-l_j (l_1\cdots {\bar l}_j\cdots  
l_n)}$ satisfy all relations expected for moments of a conserved current. We may 
thus define a ten-dimensional spacetime tensor
\[
T^{\mu \nu}(x) =  \int { d^9k \over (2 \pi )^9}
\left[e^{-ik_ix_i}\sum_n\sum_{l_1..l_n}\itt^{\mu \nu (l_1 \cdots l_n)} 
{(i)^n \over n!} 
k_{l_1} \cdots k_{l_n} \right]
\]
interpreted as the eleven-dimensional stress-energy tensor integrated over the 
compact circle. The relations we have proven show that these ten-dimensional 
currents are conserved, $\partial_{\mu}T^{\mu \nu}(x) = 0$ where $\mu$ runs 
over $\{+,1,2,...,9\}$ and $\nu = 1,2,...,9,+,$ or $-$. Note that the moments we have 
been discussing contain no information about the distribution of matter in the 
compact direction. This is encoded in the $m>0$ moments in equation 
(\ref{fourier}), and 
these did not appear in the supergravity potential arising from the exchange of 
quanta 
with no longitudinal momentum. 

\subsection{Membrane current}

The membrane current is a totally antisymmetric tensor whose spatial moments 
(averaged over the compact circle) for a given Matrix theory background were 
deduced to be

\beas
\label{eq:MatrixJ}
\ijj^{-ij(l_1\cdots l_n)} & = & {1 \over 6 R} \str \left( (\dx^i \dx^k F^{kj} -
\dx^j \dx^k F^{ki} - \half \dx^k \dx^k F^{ij} \right.\\ 
& & \qquad \qquad  \left. + {1 \over 4} F^{ij} F^{kl} F^{kl} +F^{ik}
F^{kl} F^{lj} ){\bf X}^{l_1}\cdots {\bf X}^{l_n}\right)  \\ 
\ijj^{+-i(l_1\cdots l_n)} & = & {1 \over 6 R} \str \left( F^{ij} \dx^j {\bf 
X}^{l_1}\cdots 
{\bf X}^{l_n}\right)  \\
\ijj^{ijk(l_1\cdots l_n)} & = & - {1 \over 6 R} \str \left( (\dx^i F^{jk} + \dx^j F^{ki} 
+ 
\dx^k F^{ij} ){\bf X}^{l_1}\cdots {\bf X}^{l_n}\right)  \\
\ijj^{+ij(l_1\cdots l_n)} & = & - {1 \over 6 R} \str\left( F^{ij}{\bf X}^{l_1}\cdots 
{\bf 
X}^{l_n}\right)  
\eeas
Conservation of the membrane current requires the following relations between 
the moments (\ref{zero_eqns}):
\[
\partial_t \ijj^{+\mu \nu(l_1\cdots l_n)} = \ijj^{\mu \nu l_1(l_2\cdots l_n)} + 
\cdots + 
\ijj^{\mu \nu l_n (l_1\cdots l_{n-1})}
\]
Since $\ijj$ is an antisymmetric tensor, it suffices to give a proof for $(\mu,\nu) 
= (i,j)$ and $(\mu, \nu) = (-,i)$. 

In the first case, we have 
\beas
\sum_{k=1}^n\ijj^{ijl_k (l_1\cdots {\bar l}_k\cdots l_n)} &=& -{1 \over 6R} 
\sum_{k=1}^n 
\;\str\left(F^{ij}{\bf X}^{l_1} \cdots {\bf X}^{l_{k-1}}\dx^{l_k} {\bf 
X}^{l_{k+1}}\cdots {\bf X}^{l_n} \right.\\
& &\hspace{1.3in} \left. + \; i\dx^i{\bf X}^{l_1}\cdots{\bf X}^{l_{k-1}}[{\bf 
X}^j,{\bf X}^{l_k}] {\bf X}^{l_{k+1}}\cdots 
{\bf X}^{l_n} \right.\\
& &\hspace{1.3in} \left.- \; i\dx^j{\bf X}^{l_1}\cdots{\bf X}^{l_{k-1}}[{\bf X}^i,{\bf 
X}^{l_k}] {\bf X}^{l_{k+1}}\cdots {\bf 
X}^{l_n}\right)\\
&=& - {1 \over 6R} \;\str\left(F^{ij} \partial_t({\bf X}^{l_1}\cdots {\bf X}^{l_n}) + 
(i[\dx^i,{\bf X}^j]+i[{\bf X}^i,\dx^j]){\bf X}^{l_1}\cdots {\bf X}^{l_n}\right)\\
&=&-{1 \over 6R} \partial_t \;\str(F^{ij}{\bf X}^{l_1}\cdots {\bf X}^{l_n})\\
&=&\partial_t \ijj^{+ij(l_1\cdots l_n)}
\eeas
where we have used equation (\ref{id1}) from the appendix to arrive at the 
second line.

For $(\mu, \nu) = (-,i)$, we have
\beas
\lefteqn {\sum_{j=1}^n\ijj^{-il_j (l_1\cdots {\bar l}_j\cdots l_n)}}\\
&=& {1\over 6R} \sum_{j=1}^n \;\str \left(F^{ik} \dx^k {\bf X}^{l_1}\cdots{\bf 
X}^{l_{j-1}}\dx^{l_j}{\bf X}^{l_{j+1}} \cdots 
{\bf X}^{l_n} \right.\\
& &+ \; i\dx^i \dx^k {\bf X}^{l_1}\cdots{\bf X}^{l_{j-1}}[{\bf 
X}^k,{\bf X}^{l_j}] {\bf X}^{l_{j+1}}
\cdots {\bf X}^{l_n} - {i \over 2}\dx^k\dx^k {\bf X}^{l_1}\cdots{\bf 
X}^{l_{j-1}}[{\bf X}^i,{\bf 
X}^{l_j}]{\bf X}^{l_{j+1}} \cdots {\bf X}^{l_n}\\
& &\left. + \; iF^{ik}F^{kl}{\bf X}^{l_1}\cdots{\bf X}^{l_{j-1}}[{\bf 
X}^l,{\bf 
X}^{l_j}]{\bf X}^{l_{j+1}} \cdots {\bf X}^{l_n} + {i \over 4} F^{kl} F^{kl}{\bf 
X}^{l_1}\cdots{\bf X}^{l_{j-1}}[{\bf 
X}^i,{\bf X}^{l_j}]{\bf X}^{l_{j+1}} \cdots {\bf X}^{l_n} \right)\\
&=&  {1\over 6R} \;\str \left(F^{ik} \dx^k \partial_t ({\bf X}^{l_1}\cdots {\bf 
X}^{l_n}) + \left(i[\dx^i,{\bf X}^k]\dx^k +i[{\bf X}^i,\dx^k]\dx^k\right){\bf 
X}^{l_1}\cdots {\bf X}^{l_n}\right.\\
& &\hspace{1in} \left. + \left(i[F^{kl},{\bf X}^l]F^{ik} +i[F^{ik},{\bf X}^l]F^{kl} + {i 
\over 2} [F^{kl},{\bf X}^i]F^{kl}\right){\bf X}^{l_1}\cdots {\bf X}^{l_n}\right)
\eeas
Here, we have used four applications of equation (\ref{id2}), plus the Gauss law 
constraint $[{\bf  X}^i,\dx^i] = 0$. Now, we note that 

\beas
\lefteqn {\str\left(\left(i[F^{ik},{\bf X}^l] + {i \over 2} [F^{kl},{\bf 
X}^i]\right)F^{kl}{\bf 
X}^{l_1}\cdots {\bf X}^{l_n}\right)}\\
&=& -\str\left({1 \over 2} \left[{\bf X}^k{\bf X}^l{\bf X}^i+{\bf X}^l{\bf X}^k{\bf 
X}^i+{\bf X}^i{\bf X}^k{\bf X}^l+{\bf X}^i{\bf X}^l{\bf X}^k-2{\bf X}^l{\bf X}^i{\bf 
X}^k-2{\bf X}^k{\bf X}^i{\bf X}^l\right]F^{kl}{\bf X}^{l_1}\cdots {\bf 
X}^{l_n}\right)\\
&=&0,
\eeas
since the inner expression in brackets, treated as a unit in the symmetrization,  
is symmetric in $k$ and $l$ while $F^{kl}$ 
is antisymmetric. Hence,
\beas
\lefteqn {\sum_{j=1}^n\ijj^{-il_j (l_1\cdots {\bar l}_j\cdots l_n)}}\\
&=& {1 \over 6R} \;\str \left(F^{ik} \dx^k \partial_t ({\bf X}^{l_1}\cdots {\bf 
X}^{l_n}) + \dot{F}^{ik}\dx^k{\bf X}^{l_1}\cdots {\bf X}^{l_n} + F^{ik}\ddot{X}^k{\bf 
X}^{l_1}\cdots {\bf X}^{l_n}\right)\\
&=& {1 \over 6R} \partial_t \;\str \left(F^{ik}\dx^k{\bf X}^{l_1}\cdots {\bf 
X}^{l_n}\right)\\
&=&\partial_t\ijj^{+ - i (l_1\cdots l_n)},
\eeas
as desired. In \cite{Dan-Wati2}, it was noted that the terms 
\[
\frac{1}{6 R}
  \str \left({1 \over 2} F^{ij} F^{kl} F^{kl} + F^{ik} F^{kl} F^{lj} \right).
\]
appeared in the matrix expression for $\ijj^{-ij}$ but vanished for a classical 
membrane. Their physical interpretation was therefore unclear, however, we see 
here that they are necessary for current conservation at finite N.

Thus, we have shown that the matrix theory quantities $\ijj^{\mu \nu \lambda 
(l_1\cdots l_n)}$ satisfy all relations expected for moments of a conserved 
membrane current. As with the stress energy tensor, we may define a 
ten-dimensional tensor
\[
J^{\mu \nu \lambda}(x) = \int {d^9k \over (2 \pi)^9}
\left[e^{-ik_ix_i}\sum_n\sum_{l_1..l_n}\ijj^{\mu \nu \lambda (l_1 \cdots l_n)} 
{(i)^n \over n!} k_{l_1} \cdots k_{l_n} \right]
\]
interpreted as the eleven-dimensional membrane current integrated over the 
compact circle. The matrix identities we have shown amount to a proof of the 
conservation of these ten-dimensional currents, $\partial_{\mu}J^{\mu \nu 
\lambda}(x) = 0$ where $\mu$ runs over $\{+,1,2,...,9\}$ and $\nu, \lambda = 
1,2,...,9,+,$ or $-$. Again, these moments do not give any information about the 
charge distribution in the compact direction.
 
\subsection{Fivebrane current}

In the case of the fivebrane current, not all components had moments appearing 
in the potential considered in \cite{Dan-Wati2}, so the authors were only able to 
determine expressions for the moments of the static longitudinally wrapped 
fivebrane charge, 
\[
 \imm^{+-ijkl(l_1\cdots l_n)} = {1 \over 12 R} \str \left((F^{ij} F^{kl} + F^{ik}
F^{lj} + F^{il} F^{jk}){\bf X}^{l_1}\cdots {\bf X}^{l_n}\right)\,.
\]
However, these moments appear together with moments of $\imm^{-ijklm}$ in 
the relations imposed by conservation of fivebrane current. Requiring these 
relations to be satisfied in Matrix theory we are led to define
\bea
\imm^{-ijklm(l_1\cdots l_n)} &=& {1 \over 12R} \;\str\left(\left(
\dx^i(F^{jk}F^{lm} + F^{jl}F^{mk} + F^{jm}F^{kl})\right. \right.\label{fivecurrent}\\
& & \hspace{0.7in} \;+\;\dx^j(F^{kl}F^{mi} + F^{km}F^{il} + 
F^{ki}F^{lm})\nonumber\\
& & \hspace{0.7in} \;+\;\dx^k(F^{lm}F^{ij} + F^{li}F^{jm} + 
F^{lj}F^{mi})\nonumber\\
& & \hspace{0.7in} \;+\;\dx^l(F^{mi}F^{jk} + F^{mj}F^{ki} + 
F^{mk}F^{ij})\nonumber\\
& & \hspace{0.7in} \left. \left. +\;\dx^m(F^{ij}F^{kl} + F^{ik}F^{ij} + F^{il}F^{jk}) 
\right) 
{\bf X}^{l_1}\cdots {\bf X}^{l_n} \right)\nonumber
\eea
With this definition, it may be checked using equation (\ref{id2}) that the 
conservation relation 
\[
\partial_t\imm^{+ - ijkl(l_1\cdots l_n)} = \imm^{- ijkl l_1(l_2\cdots l_n)} + 
\cdots + 
\imm^{- ijkl l_n (l_1\cdots l_{n-1})}
\]
is satisfied. The remaining components of the totally 
antisymmetric membrane current, $\imm^{+ijklm}$ and $\imm^{ijklmn}$ are 
the charge and spatial current of the transverse fivebrane, poorly understood 
and perhaps vanishing in Matrix theory.  

\section{Discussion}

The results described here provide further evidence for the connection between 
Matrix 
theory and eleven-dimensional supergravity. Starting with the Matrix theory 
quantities identified in \cite{Dan-Wati2} as corresponding to moments of the 
stress-energy 
tensor and membrane current, we have demonstrated that all relations required 
by current conservation hold as exact matrix identities at finite N. For the 
longitudinal fivebrane current, only the static charge had been identified 
previously in \cite{Dan-Wati2}. By requiring consistency with the fivebrane 
current 
conservation equations, we have been led to a definition for moments of the 
spatial current associated with this charge (equation (\ref{fivecurrent})). 

Our results show that for a given Matrix theory background, we may associate 
corresponding ten-dimensional stress-energy, membrane, and longitudinal 
fivebrane currents (the eleven-dimensional currents averaged around the 
compact circle) which are conserved in the ten-dimensional sense.  This current 
conservation, manifested in Matrix theory as the series of non-trivial identities 
we have proven, is essential to the validity of the correspondence with 
supergravity. 

Though convincing evidence has been given that finite-N Matrix theory is a 
description of DLCQ M theory \cite{Sen,Seiberg-DLCQ}, it is not clear that DLCQ 
M-theory has a low energy description as DLCQ supergravity (see however, 
\cite{bgl}).  Many properties of supergravity, including many aspects of the long 
range interactions between objects, seem to be reproduced correctly in finite N 
matrix theory,  however others, such as the equivalence principle 
\cite{Dan-Wati2} seem only to be recovered in the large-N limit of matrix theory. 
Thus, it is particularily interesting that our results hold exactly at finite N.

It is important to note that the moments we have been dealing with 
contain no information about the distribution of matter or charge in the 
longitudinal direction. In supergravity, this information appears as non-zero 
fourier modes of the current around the compact circle ($m > 0$ in equation 
(\ref{fourier})). 
These moments do not contribute to the supergravity potential arising from the 
exchange of quanta with 
zero longitudinal momenta, so if corresponding Matrix theory quantities were to 
be deduced in an analogous way it would probably require a study of processes 
involving longitudinal momentum transfer. Actually, it is unclear to what extent 
a distribution of matter and charge in the compact direction is encoded in the 
finite-N Matrix theory variables.  If DLCQ supergravity is not an appropriate 
description of DLCQ M-theory, then it 
may be incorrect to speak classically of a full eleven-dimensional current. In 
certain sectors of Matrix theory, for 
example matrices which describe membranes with no winding around the 
compact direction, a classical current distribution in the longitudinal direction 
(up to an undefined constant) certainly is encoded in Matrix theory, at least in 
the large N limit. However, there may be some ambiguity at finite N, in particular 
for states with winding around the compact circle. 


\section*{Acknowledgements}

I am grateful to Wati Taylor for helpful advice and many useful discussions. I 
would also like to thank Dan Kabat for helpful comments. This work was 
supported by an NSERC PGS B fellowship.

\section*{Appendix}

In this appendix, we prove the following matrix identity, useful in demonstrating 
the Matrix theory relations implied by current conservation. Let $A_i, B_j,$ and 
$C$ be $N \times N$ matrices with $i \in \{1,...,n\}$ and $j \in \{1,...,m\}$. Then

\be
\sum_{j=1}^m\str \left( A_1\cdots A_nB_1\cdots B_{j-1}[C,B_j]B_{j+1}\cdots 
B_m\right) = 
\sum_{i=1}^n \str \left(A_1\cdots A_{i-1}[A_i,C]A_{i+1}\cdots A_nB_1\cdots 
B_m\right)
\label{identity}
\ee
where the commutator is to be treated as a unit in the symmetrization. 

The proof is straightforward. Consider the set of terms on the left side of 
(\ref{identity}) with 
a particular ordering of the $A$'s and $B$'s. Noting the simple identity
\[
\sum_i B_{l_1}\cdots B_{l_{i-1}}[C,B_{l_i}]B_{l_{i+1}}\cdots B_{l_k} = 
CB_{l_1}\cdots B_{l_k} - B_{l_1}\cdots B_{l_k}C\,,
\]
we see that upon expansion of the commutators, all terms with $C$ sandwiched 
between two $B$'s will cancel. The resulting expression is the sum of all 
possible insertions of $C$ between an $A$ and a $B$ in the sequence of $A$'s 
and $B$'s in the original  order, where terms containing $A_iCB_j$ appear with a 
positive sign, and terms containing $B_jCA_i$ appear with a negative sign.  
Applying the same reasoning to the set of terms on the right side of 
(\ref{identity}) with this 
same ordering of $A$'s and $B$'s, we arrive at an identical expression. Since we 
have considered an arbitrary ordering of $A$'s and $B$'s, the proof of 
(\ref{identity}) is 
complete, as the symmetrized trace is just the average of all such orderings.

In deriving the Matrix theory current conservation relations, we will use the 
following special cases of this identity. First, taking n=1 above, we have
\be
\sum_{i=1}^n\str \left( AX^{l_1}\cdots X^{l_{i-1}}[C,X^{l_i}]X^{l_{i+1}}
\cdots X^{l_n}\right) = \str 
\left([A,C]X^{l_1}\cdots X^{l_n}\right)
\label{id1}
\ee
Taking n=2 above, we have
\be
\sum_{i=1}^n\str \left( ABX^{l_1}\cdots X^{l_{i-1}}[C,X^{l_i}]X^{l_{i+1}}\cdots 
X^{l_n}\right) = \str 
\left(([A,C]B+[B,C]A)X^{l_1}\cdots X^{l_n}\right)
\label{id2}
\ee
As above, the commutator is to be treated as a unit in the symmetrization for 
these equations. 

\bibliographystyle{plain}

\begin{thebibliography}{10}

\bibitem{BFSS}
T.\ Banks, W.\ Fischler, S.\ Shenker, and L.\ Susskind, ``M Theory as a Matrix
  Model: A Conjecture,'' {\tt hep-th/9610043}.

\bibitem{Susskind-review}
D.\ Bigatti and L.\ Susskind, ``Review of Matrix theory,'' {\tt
  hep-th/9712072}.
 
 \bibitem{BanksReview}
T.\ Banks, ``Matrix Theory,'' {\tt hep-th/9710231}.
  
  \bibitem{Wati-review}
  W.\ Taylor, ``Lectures on D-branes, Gauge Theory, and M(atrices),'' {\tt 
hep-th/9801182}.
  
  \bibitem{Dan-Wati}
D.\ Kabat and W.\ Taylor, ``Spherical membranes in Matrix theory,'' {\tt
  hep-th/9711078}.

\bibitem{Mark-Wati}
W.\ Taylor and M.\ Van Raamsdonk, ``Angular momentum and long-range
  gravitational interactions in Matrix theory,'' {\tt hep-th/9712159}.

\bibitem{Dan-Wati2}
D.\ Kabat and W.\ Taylor, ``Linearized supergravity from Matrix theory,'' {\tt
  hep-th/9712185}.

\bibitem{Sen}
A.\ Sen, ``D0 Branes on $T^n$ and Matrix Theory,'' {\tt hep-th/9709220}.

\bibitem{Seiberg-DLCQ}
N.\ Seiberg, ``Why is the Matrix Model Correct?,'' \PRL {\bf 79} (1997) 3577,
  {\tt hep-th/9710009}.

\bibitem{Dine-Rajaraman}
M.\ Dine and A.\ Rajaraman, ``Multigraviton Scattering in the Matrix Model,''
  {\tt hep-th/9710174}.

\bibitem{Douglas-Ooguri}
M.\ Douglas and H.\ Ooguri, ``Why Matrix Theory is Hard,'' {\tt
  hep-th/9710178}.
  
\bibitem{bgl}
V.\ Balasubramanian, R.\ Gopakumar and F.\ Larsen, ``Gauge theory, geometry and
  the large N limit,'' {\tt hep-th/9712077}.

\bibitem{Lowe}
D.\ Lowe, ``Statistical Origin of Black Hole Entropy,'' {\tt hep-th/9802173}.

\end{thebibliography}

\end{document}